\documentstyle[aps]{revtex}
\def\be{\begin{equation}}
\def\ee{\end{equation}}
\begin{document}
\draft
\title{ \Large
Effect of  the Coulomb repulsion on the {\it ac} transport through a
quantum dot
        }
\author{
 \large   T. Ivanov$^{(a)}$, V. Valtchinov$^{(b)}$,
and L. T. Wille$ ^{(c)}$\\
        }
\address{
$^{(a)}$ Department of Physics,
University of Sofia, 5 J. Baucher Blvd, 1126 Sofia, Bulgaria\\
$^{(b)}$ Department of Physics, Norteastern University, Boston, MA
 02115, USA\\
$^{(c)}$ Department of Physics, Florida Atlantic University, Boca Raton,
FL 33431-0991\\
                }
\maketitle
\begin{abstract}
We calculate in a linear response the admittance of a quantum dot out of
equilibrium. The interaction between two electrons with
opposite spins simultaneously residing on
the resonant level is modeled by an Anderson Hamiltonian. The electron
correlations lead to the appearence of a new feature in the frequency
dependence of the conductance. For certain parameter values there are two
crossover frequencies between
a capacitive and an inductive behavior of the imaginary part of
the admittance. The experimental implications of the obtained results are
briefly discussed.
\end{abstract}
The double-barrier resonant tunneling systems (DRBTS) are in the focus
of intensive
experimental \cite{1} and theoretical
\cite{beenakker91,gro91,cheng91,but86,her91,glaz88,ng88} investigations.
While the main characteristics of the steady-state transport properties of
such systems have been well inderstood in terms of several basic
approaches -- the kinetic equation approach
\cite{glaz88,ng88}, by a Landauer-B\"{u}ttiker-type formula \cite{but86},
and via the Wigner-function
\cite{kluksdahl} -- some of the features in the time-dependent behavior
of the DRBTS still remain unclear. In particular, the long lasting question
of the fact whether the tunneling transport through the structure is via an
establishing a coherent wave function, or can be viewed as a sequence of
quantum tunneling events \cite{gering87,brown89}. The answer to this question
is of major importance to the practical applicability of these devices as a
high frequency resonators \cite{frensley86,chen90,fu93,ivanov93}.

In a recent paper Fu and Dudley \cite{fu93} have utilized
a model of non-interacting electrons transmitted through the structure under
the
influence of a small {\it ac} bias superimposed upon the driving {\it dc}
voltage, having shown that for some values of the system's parameters the
frequency-driven behavior can be simulated by an equivalent electrical
circuit with an
additional inductive element -- feature, which was earlier encountered in the
modelling the frequency dependence of the DRBTS (\cite{gering87,brown89}).
In this context, however, it is important to correctly account for
the Coulomb interaction
effects due to the repulsion between two electrons simultaneously residing on
the resonant level -- phenomenon, which was shown to give rise of a numerous
effects in the {\it dc} transport \cite{gro91,her91,glaz88}. In a previous
paper \cite{ivanov93} we have calculated the
linear response
admittance of a quantum dot with interacting electrons in the case of a zero
{\it dc} voltage, thus isolating the effects due to the Coulomb on-site
repusion in an equilibrium state of the system. We have obtained a feature in
both the real and the imaginary part of the admittance at $\Omega \sim E_c$. In
a recent paper
Brandes, Weinmann, and Kramer \cite{brandes93} have studied the {\it ac}
conductance of
a tunnel junction in a linear response approximation. Having introduced the
picture of a simultaneous transmission of a electron-hole pair through the
system in the {\it ac} transport regime, they argued that
at high
frequencies the Coulomb interaction would be of minor importance since the
pair is electro-neutral. They estimated the upper frequency limit for the
Coulomb interaction to be noticeable $\omega_0 \approx 1 GHz$ for the
currently accsessible experimental parameters.

In this work we apply a non-equilibrium technique to this problem to deal
with the case of a non-zero applied {\it dc} voltage. We study the effects of
the Coulomb interaction on the relatively low-frequency behavior of the
conductance through the dot. It is worth mentioning at this point
that the formalism developed here is equally applicable either to the case
of symmetrical ($T_L=T_R$) or asymmetrical ($T_L \ne T_R$) coupling of
the well to the leads \cite{jacob93}, and can be easily extended to take into
account additional effects due to environmental fluctuations
(i.e. electron-phonon coupling).
The conductance is calculated as a function
of the
 {\it dc} voltage and the frequency $\Omega$ of the applied {\it ac} bias for
$\Omega$ of the order of several $\gamma$, where $\gamma$ is the elastic
width of the resonant level.

We found that the Coulomb repulsion has a profound
effect on both the frequency-driven conductance and the energy losses over a
relatively broad range of parameters. In particular, a new feature in the
{\it ac} response is found in the case of finite Coulomb repulsion energy
$E_c$, which we contribute to the fundamental way in which the well-electrons
energy spectrum cnanges in the presence of electron-electron interactions -
and this is in marked contrast to the non-interacting $(E_c \to 0)$ limit.
The imaginary part of
the admittance behaves in a way consistent with the Kramers-Kronig dispersion
relations, thus giving rise to an additional featute for a finite $E_c$.
We further discuss in some details the experimental implications of the
theoretical findings reported here,  giving the conditions
necessary for observing the predicted effects, and discussing some recent
experimental results as well.

The Hamiltonian of the quantum well, coupled to the leads, can be written as
\FL
\begin{equation}
H = \sum\limits_{k\sigma}\epsilon_{k}^{L}a^{\dag}_
{k\sigma}a_{k\sigma}+
\sum\limits_{p\sigma}\epsilon_{p}^{R}b^{\dag}_{p\sigma}b_{p\sigma} +
\epsilon_{c} \sum_{\sigma}c_{\sigma}^{\dag}c_{\sigma}+
E_{c}n_{\uparrow}n_{\downarrow} +
\sum\limits_{k\sigma}(T_{Lk}c^{\dag}_{\sigma}a_{k\sigma}+h.c)+
\sum\limits_{p\sigma}(T_{Rp}b^{\dag}_{p\sigma}c_{\sigma} + h.c).
\end{equation}
It is expressed in terms of the creation
(annihilation) operators in the emitter $a_{k\sigma}$, collector
$b_{p\sigma}$,
and the well ($c_{\sigma}$) with $k(p)$ the corresponding quasimomenta
and $\sigma=\uparrow, \downarrow$ the spin index.
Here $\epsilon^{L}_{k}$ and $\epsilon^{R}_{p}$ are the single-particle
energies in the emitter and the collector leads, respectively. $E_{c}$ is
the Coulomb repulsion between electrons with opposite spins. The
single-electron energies are measured from the corresponding Fermi levels
$\mu_{L}$ and $\mu_{R}$ in the emitter and collector, and the $dc$
bias is $\mu_{L}-\mu_{R}=eV$. $\epsilon_c=\epsilon^0_c+ \alpha eV$, where
$\epsilon^0_c$ is the bare resonant level energy and $\alpha$ measures the
portion of the voltage drop on the quantum well($\alpha \approx 0.5$ for
the symmetric structure $T_{Lk}=T_{Rp}$). The particle-number operator is
$n_{\sigma}=c^{\dag}_{\sigma}c_{\sigma}$.
$T_{Lk}$ and
$T_{Rp}$ are the tunneling matrix elements which are bias dependent.

In order to investigate the response of the system to an external
perturbation
we introduce an additional $ac$ voltage, which is superimposed on the $dc$
bias $eV$. The Hamiltonian of this $ac$ signal has the form:
\FL
\begin{equation}
H_{ext}(t)=-\alpha eu(t)\sum\limits_{\sigma} c_{\sigma}^{\dag}c_{\sigma} -
eu(t)\sum\limits_{p\sigma}b^{\dag}_{p\sigma}b_{p\sigma} + h.c.,
\label{pert}
\end{equation}
where the field operators are taken at the time $t$.  The external field is
$u(t)=u_{0}e^{i\Omega t}$
and its amplitude  is assumed to be very small in order not to disturb the
electron distribution in the well. In writing this form of the $H_{ext}$
we have set $\mu_{R}=0$ \cite{chen90}.

First we calculate the electron Green's functions.
We apply the non-equilibrium
(Keldysh) formalism where one introduces the retarded(advanced) and
distribution Green's functions \cite{keldysh,zhou}. We assume that the
relaxation
processes in the leads are much faster than in the quantum well. Thus, we can
consider the leads as equilibrium systems and the corresponding Green's
functions are given by the usual expressions for a non-interacting electron
system \cite{keldysh,zhou}.

The retarded (advanced) $G_{r(a)}(\omega)$ and the
distribution
$G_{<}(\omega)$ Green's
function for the electrons
in the well corresponding to the full Hamiltonian $H$ are determined
using the irreducible Green's function method \cite{tyablikov}. This method
has been applied successfully to the Hubbard model \cite{kuz89}.  Its main
advantage
consists of treating all the truncations of the higher order
Green's functions
with the same accuracy -- consistent with already
choosen algebra of relevant to the problem operators. It is worth
mentioning that it is valid both in the weak and the strong correlation
limits.

We recall that in the non-equilibrium formalism the Green's functions
depend on two
time variables
 $t_{1}, t_{2}$. It can be shown that the retarded (advanced) Green's
functions depend only on the "relative" time $t = t_1 - t_2$ while the
distribution Green's function depends on both $t$ and
$T=\frac{t_1+t_2}{2}$ \cite{zhou}.

The well electrons Green's function is obtained in the form (for more details
see \cite{ivanov93})
\FL
\begin{equation}
G_r (\omega) = {\omega - \tilde {\epsilon}_{c\sigma} -
\Sigma_0 - \Sigma_1 \over {(\omega - \epsilon_{c} - \Sigma_0)(\omega -
\epsilon_c - \Sigma_0 - \Sigma_1) - E_c (\omega - \epsilon_c - \Sigma_0 -
\Sigma_2) }}.
\label{general}
\end{equation}
The explicit expressions for the self-energy parts $\Sigma_0, \Sigma_1,
\Sigma_2$ in Eqn.~(\ref{general}) can be found in Ref. 15. We use the
following
notation: $\tilde{\epsilon}_{c\sigma}=\epsilon_{c} + E_{c}(1-<n_{-\sigma}>)$
where $<n_{\sigma}>$ is the average number of well
electrons with spin $\sigma$.

This Green's function describes two energy levels for the quantum well
electrons - a lower level with energy $\epsilon_c$ and an upper level with
energy $E_c + \epsilon_c$.

We should stress
that the derivation of the Green's function presented
here is valid for temperatures higher than the characteristic temperature
for this problem -- the Kondo temperature $T_K$.
Lacroix \cite{lacroix} has
shown that for temperatures $T<T_K$ this truncation
procedure omits terms which are divergent at the Fermi level. These terms
give rise to the Kondo effect.

The distribution Green's function is
calculated assuming that all transient processes after the
switching on the $dc$ bias have decayed. In this case one can consider the
distribution Green's function to be independent on the time $T$.
The result for $G_{<}(\omega)$ is found in the following closed form:
\FL
\begin{equation}
G_< (\omega)= - F(\omega) \left ( G_r (\omega) - G_a (\omega) \right )
\label{distr}
\end{equation}
where
\FL
\begin{equation}
F(\omega)=  \frac{ \sum \limits_{k}
 \left |T_{Lk} \right |^{2}A_{<}
(k, \omega) + \sum \limits_{p} \left |T_{Rp} \right |^{2}B_{<}(p, \omega)}
{\sum\limits_{k}
\left |T_{Lk} \right |^{2}(A_{r}(k, \omega)-A_{a}(k, \omega))
+\sum\limits_{p} \left |T_{Rp} \right |^{2}
(B_{r}(p, \omega) - B_{a}(p, \omega) ) }
\end{equation}
is the new non-equilibrium (but steady-state) distribution of electrons
in the
quantum well. In this expression $A_{r(a)} \ (B_{r(a)})$ and
$A_< \ (B_<)$ are
the retarded (advanced) and the distribution Green's functions in the left
(right) lead, respectively.

Next we calculate the current through the quantum well in the presence of a
time-dependent electric field.
The total current is
 given by $I=(I_{L}+I_{R})/2$ where $I_{L(R)}$ is the current through the
 left(right) barrier, respectively
\FL
\begin{eqnarray}
I(t)=-\frac{ie}{2} \langle \sum\limits_{k\sigma}
\left [ T_{Lk}c^{\dag}_{\sigma}(t)
     a_{k\sigma}(t)-T^{\ast}_{Lk}a^{\dag}_{k\sigma}(t)c_{\sigma}(t)\right]
\nonumber
\\
  +\sum\limits_{p\sigma} \left [T_{Rp}b^{\dag}_{p\sigma}(t)c_{\sigma}(t)-
 T^{\ast}_{Rp}
     c^{\dag}_{\sigma}(t)b_{p\sigma}(t) \right ] \rangle.
\label{comut}
\end{eqnarray}
To obtain this expression we have used the Shockley - Ramo theorem
\cite{shoram}.

In order to calculate the current $i(t)$ due to the external $ac$
voltage $u(t)$, we use the linear response formalism.
In this way we obtain the admittance of the
quantum dot $Y(\omega) = i(\omega)/u(\omega)$ where $u(\omega) = u_0
(\delta(\omega - \Omega) + \delta(\omega + \Omega))$ is the Fourier transform
of the external $ac$ bias. The explicit expression for the admittance is
obtained in the form:
\FL
\begin{eqnarray}
Y(\Omega)& =& \int d\omega \left\{
- 2 i \left( \alpha + D_r (\omega ,
\Omega)
\right) \left( \gamma_L (\omega) f_L (\omega) - \gamma_R
(\omega) f_R (\omega) \right)  G_r (\omega + \Omega) G_r
(\omega)   \right.  \nonumber \\
&+&\left. 2i \left( \alpha + D_a (\omega ,
\Omega)
\right) \left( \gamma_L (\omega + \Omega) f_L (\omega + \Omega) - \gamma_R
(\omega + \Omega) f_R (\omega + \Omega) \right)  G_a (\omega + \Omega) G_a
(\omega)   \right.  \nonumber \\
&-&\left. \left [ \left( \alpha + D_a (\omega + \Omega) \right) F (\omega +
\Omega) \left( G_r (\omega + \Omega) - G_a (\omega + \Omega) \right) G_a
(\omega) \right. \right.  \\
&+&\left. \left. \left( \alpha + D_r (\omega + \Omega) \right) F (\omega)
G_r (\omega + \Omega) \left( G_r (\omega) - G_a (\omega)
\right)  \right] \right. \nonumber \\
&\times & \left. \left( A_r (\omega + \Omega) - A_a (\omega) - B_r (\omega +
\Omega) + B_a (\omega) \right) \right\}
\label{admit}
\end{eqnarray}

In writing Eqn. ~(\ref{admit}) we have used the following notations:
\FL
\be
A_{r(a)} (\omega) = \sum \limits_k \left |T_{Lk} \right |^2 A_{r(a)} (k,
\omega),
\ee
\FL
\be
B_{r(a)} (\omega) = \sum \limits_p \left |T_{Rp} \right |^2 B_{r(a)} (p,
\omega),
\ee
\FL
\be
D_{r(a)} (\omega, \Omega) = \sum \limits_p \left |T_{Rp} \right |^2 B_{r(a)}
(p, \omega + \Omega) B_{r(a)} (p, \omega).
\ee
The tunneling matrices $T_{Lk}$ and $T_{Rp}$ are
related to the level widths for the leads' electrons via the usual equations
$
\gamma_L(\omega)= \pi \sum\limits_{k} \left |T_{Lk} \right |^{2}
\delta(\omega-\epsilon^{L}_{k}),
\gamma_R(\omega)= \pi \sum\limits_{p} \left |T_{Rp} \right |^{2}
\delta(\omega -  \epsilon^{R}_{p})
$. In the following $\gamma_L(\omega), \gamma_R(\omega)$ are taken to be
independent of $\omega$.

Now we present our numerical results for the $ac$ conductance $\sigma(\Omega) =
Re Y(\Omega)$ and the energy losses $Im Y(\Omega)$ through the quantum well.
We calculate them for $T > T_K$ assuming
that $E_c \gg T, \gamma=\gamma_L+\gamma_R$ and $T \gg \gamma$. First we solve
self-consistently the equation for the average number of quantum well electrons
$\langle n_{\sigma} \rangle = \langle n_{-\sigma} \rangle = n =
-\int d\omega/2\pi Im G_{<}(\omega)$ (we consider a non-magnetic solution). We
take a broad flat density of states for the leads' electrons.
In Fig. 1 we show the dynamical conductance $\sigma(\Omega, V)$ calculated for
a bare level energy $\epsilon^{(0)}_c = 0.2 E_c$ and for a symmetrical coupling
of the leads to the well - $\gamma_L = \gamma_R$.

For $\Omega \to 0$ ($dc$ limit) $\sigma(V)$ has two maximums. This structure
reflects the energy spectrum of the well electrons - there are two channels
for
the electrons to tunnel through the well. When the number of well electrons is
smaller than one ($n < 0.5$) the tunneling is predominantly through the lower
(resonant) level. For $n > 0.5$ the lower level is filled with electrons and
they are transfered through the upper level.

In this paper we show that the effect of the electron correlations
(the Coulomb
repulsion $E_c$) can also be observed in the $\Omega$ dependence of the
conductance. For relatively low frequency $(\Omega \sim 2-3 \gamma)$ the
conductance decreases with $\Omega$ (similarly to the case of non-interacting
electrons \cite{chen90,fu93}) since the electrons cannot follow the
applied {\it ac}
field. The electrons build-up in the well and fill the upper level (there is
more than one electron in the well). This opens an additional tunneling channel
through the upper level. Consequently for higher frequencies the tunneling
current increases (the feature at $\sim 8-10 \gamma$ for relatively low {\it
dc} voltage). For larger frequencies the conductance again falls off with
$\Omega$, it becomes negative and tends to zero with negative values. This
behaviour is in marked contrast to the non-interacting case. In the latter the
conductance is positive (for a resonant level above the right chemical
potential) and never changes sign.

When the {\it dc} voltage increases the feature in the $\Omega$ dependence of
$\sigma$ we have just discussed is almost smeared out. The conductance is a
monotonically decreasing function of $\Omega$ but the non-zero
response to the {\it ac}
field spreads to substantially higher frequencies compared to the
non-interacting case.

For {\it dc} voltages in the region of the second peak (at $\Omega \to 0$)
when
the renormalized level is brought well above $\mu_R$ the conductance is a
non-monotonic function of $\Omega$. It slightly increases for $\Omega < 5
\gamma$ and then decreases with non-zero values of $\sigma(\Omega)$ up to
$\Omega \sim 15 \gamma$.

In Fig. 2 we show the imaginary part of the admittance $Im \ Y(\Omega)$
calculated for two typical cases: a) $\epsilon^{(0)}_c = -0.112 E_c$ and b)
$\epsilon^{(0)}_c = -0.01 E_c$. The applied {\it dc} voltage was taken
to be $eV = 0.2 E_c$ and $\gamma = 0.04E_c$. When
the upper level is above the right chemical
potential the admittance shows a capacitive behavior - $Im \ Y(\Omega) >
0$( Fig. 2, curve a)).
Note that the sign of the imaginary part of the admittance is opposite to the
sign in Ref.~\cite{fu93}.

Particularly interesting is the result presented in Fig. 2, curve b). In
this case $|\epsilon_c - \mu_R| < \gamma/2$. It shows
that there are two frequencies at which a crossover from a capacitive to an
inductive behavior (and vice versa) is obtained. At low frequencies the
admittance is inductive, then it changes sign and this is the crossover to a
capacitive admittance. Fu and Dudley \cite{fu93} studied the same case
in the
non-interacting picture of the resonant tunneling. They obtained an inductive
behavior - $Im \ Y(\Omega) < 0$ and no crossover to a capacitive one.
Moreover, in the interacting case there is one more crossover frequency at
which the admittance changes back from a capacitive to an inductive. When $V =
0$ this frequency is $\Omega \sim \epsilon^{(0)}_c + E_c - \mu_R$. With
increasing $V$ the first crossover frequency diminishes and the second
increases so that at sufficiently high voltage the behaviour of the admittance
will be inductive. The appearence of the second crossover frequency is in
agreement with the conclusion of Brandes, Weinmann, and Kramer
\cite{brandes93} - at
high frequencies the behavior of the admittance is not strongly affected by
the electron interactions except the feature at $\Omega \sim E_c$
discussed in \cite{ivanov93}.

When $|\epsilon_c - \mu_R| > \gamma/2$ \ (as in Fig. 2, curve a) Fu and
Dudley
obtained a crossover to an inductive admittance at $\Omega \sim |\epsilon_c -
\mu_R|$. With $E_c \ne 0$ this is possible only if $\epsilon_c$ is well
below $\mu_R$ and the upper level is above $\mu_R$ \cite{ssc93}. For
a quantum dot with both the resonant and the upper level well below the
right chemical potential for low $V$ the admittance is inductive. A
crossover
to a capacitive behavior can be obtained for sufficiently high $V$ at
$\Omega
\sim \epsilon_c + E_c - \mu_R$ \ (then the upper level will be moved above
$\mu_R$) \cite{ssc93}.

All these results clearly indicate that as in the non-interacting case
\cite{fu93} the frequency behavior of DBRTS cannot generally be simulated
by any LRC equivalent-circuit model.

To address the experimental detection of the the effects under
consideration here,
let us recall the range of the three main parameters involved
in the model -- $E_c \gg T \gg \gamma$. The on-site repulsion energy can be
estimated from $E_c \sim e^2/\epsilon L$, where $L$ is the
size of the confined region, and $\epsilon$ is the dielectric constant for the
$GaAs$. Thus for a quantum dot with average size of 100 $\mbox{\AA}$, one gets
$E_c \sim 1 - 10 meV$. It is less evident how to estimate the elastic
broadening constant $\gamma$, but we can use the estimation given in
Ref.~\cite{foxman93} for $\gamma \sim 10 - 20 \mu eV$ for a structure of about
the same size. Therefore, the type of effects discussed in this study to be
detected experimentally, one needs temperatures of few degrees $K$ and a
external frequency range up to hundreds $GHz$ -- requirements accessible at
the present time. Let us only mention at this place that because of the
temperature ranges used ($T=77K$, and room temperature, accordingly)
the results of
two recent {\it ac} experiments
(see Refs.~\cite{gering87,brown89}) did not show the frequency dependence
discussed here. This is because the measurements were performed in the
regime
where the thermal fluctuations prevail over the ``Coulomb
blockade'' effect \cite{averin91}.

We should point out that the expression for the current Eqn.~(\ref{comut})
does not include the displacement currents through the parasitic
capacitances. They can be accounted for by considering some electrostatic
model of the dot. Our aim was to extract the effects of the electron
correlations comparing our results with the available works on
non-interacting electron tunneling where the displacement currents were
not included. One can expect that the parasitic currents would modify the
relatively high frequency behavior of the admittance \cite{fren89}.

In conclusion, we have calculated the linear-response admittance of a
quantum dot
with interacting electrons. We show that the effect of the electron
correlations can be observed in the frequency dependence of the conductance
where a new feature appears when the electrons tunnel through the upper level.
For a certain parameter values there are two frequencies at which the
imaginary part of the admittance changes sign i.e. a crossover between a
capacitive and an inductive behavior.We discuss the experimental
conditions for observation of these effects.

\acknowledgements

The authors acknowledge the fruitful discussions with Drs.
A. Groshev, S. Hershfield, V. Popov, I. Z. Kostadinov, J. Sokoloff. V.V.
is thankful to Dr. N. S. Wingreen for making papers available before
publication.
T.I. was sponsored by a Contract F-225/1992 with the Ministery of Science and
Education of Bulgaria.
\newpage

\begin{figure}
\caption {
Conductance $\sigma$ as a function of the applied
{\it dc} bias across the structure
 and the external frequency $\Omega$ calculated for a quantum dot with
 $\epsilon^{(0)}_c = 0.2 E_c $.
        }
\end{figure}
\begin{figure}
\caption{
The imaginary part of the dynamical admittance $Im Y$
as a function of the frequency $\Omega$ of the external {\it ac} signal
for two positions of the bare energy level: a)
$\epsilon^{(0)}_c = -0.112 E_c$ (the solid line), and
b) $\epsilon^{(0)}_c=-0.01 E_c$ (the dashed line). The
applied {\it dc} voltage is $eV=0.2 E_c$, and $\gamma = 0.04 E_c$.
        }
\end{figure}


\begin{references}
%
\frenchspacing
%
\bibitem{1}  M. A. Reed, J. H. Randall, R. J. Aggarwal, R. J. Matyi,
      T. M. Moore and A. E. Wetsel, Phys. Rev. Lett. {\bf 60}, 535 (1988);
 M. W. Dellow, P. H. Beton, C.J.G.M. Langerak,
  T. J. Foster, P. C. Main, L. Eaves, M. Henini, S. P.
Beaumont,
    and C. D. W. Wilkinson, Phys. Rev. Lett. {\bf 68}, 1754
(1992).
%
\bibitem{beenakker91} C. W. J. Beenakker, Phys. Rev. {\bf B 44}, 1646 (1991).
%
\bibitem{gro91} A. Groshev, T. Ivanov, and V. Valtchinov,
 Phys. Rev. Lett. {\bf 66}, 1082 (1991).
%
\bibitem{cheng91}
       L. Y. Cheng and C.S. Ting, Phys. Rev. Lett. {\bf 44}, 5916 (1991).
%
\bibitem{but86} M. B\"{u}ttiker, Phys. Rev. Lett. {\bf 57}, 1761 (1986);
        Y. Meir, N. S. Wingreen, and P. A. Lee, Phys. Rev.
      Lett. {\bf 66}, 3048 (1991).
%
\bibitem{her91}   S. Hershfield, J. Davies, and
  J. Wilkins, Phys. Rev. Lett. {\bf 67}, 3720 (1991);
     Phys. Rev. B {\bf 46}, 7046 (1992).
%
\bibitem{glaz88} L. I. Glazman and M. E. Raikh, Pis'ma Zh. Eksp. Teor. Fiz.
                    {\bf 47}, 378(1988) (JETP Lett. {\bf 47}, 452 (1988)).
%
\bibitem{ng88} T. K. Ng and P. A. Lee, Phys. Rev. Lett. {\bf 61}, 1768
(1988).
%
\bibitem{kluksdahl} N. C. Kluksdahl, A. M. Kriman, D. K. Ferry, and
C. Ringhofer, Phys. Rev. {\bf B 39}, 7720 (1989).
%
\bibitem{gering87} J. M. Gering, D. A. Crim, D. G. Morgan, P. D. Coleman,
W. Kopp, and H. Morcoc, J. Appl. Phys. {\bf 61}, 271 (1987).
%
\bibitem{brown89} E. R. Brown, C. D. Parker, and T. C. L. Sollner,
Appl. Phys. Lett. {\bf 54}, 934 (1989).
%
\bibitem{frensley86} W. R. Frensley, Phys. Rev. Lett. {\bf 57}, 2853 (1986);
Phys. Rev. B {\bf 36}, 1570 (1987).
%
\bibitem{chen90}  L.Y. Chen and C.S. Ting, Phys. Rev. Lett. {\bf
  64}, 3159 (1990); Phys. Rev. B {\bf 43}, 2097 (1991).
%
\bibitem{fu93} Y. Fu and S. C. Dudley, Phys. Rev. Lett. {\bf 70}, 65 (1993).
%
\bibitem{ivanov93} T. Ivanov, D. Marvakov, V. Valtchinov, and L. T.
                        Wille,  Phys. Rev. B {\bf 48}, 4679 (1993).
%
\bibitem{brandes93} T.Brandes, D. Weinmann, and B. Kramer,
       Europhys. Lett. {\bf 22},   51 (1993).
%
\bibitem{jacob93} C. Jacoboni, and P. J. Price, Phys. Rev. Lett.
{\bf 71}, 464 (1993).
%
\bibitem{keldysh} L.V. Keldysh, Zh. Eksp. Theor. Fiz. {\bf 47}, 1515 (1964)
      (Sov. Phys. JETP {\bf 20}, 1018 (1965)).
%
\bibitem{zhou}  K.  C. Zhou,  Z. B. Su, B. L. Hao, and Lu  u,
   Phys. Rep. {\bf 118}, 1 (1985).
%
\bibitem{tyablikov} S. B. Tyablikov, {\it Metodi kwantowoi teorii
                       magnetizma}, Nauka (Moskow, 1975).
%
\bibitem{kuz89} A. L. Kuzemsky, Doklady Akad. Nauk. SSSR, {\bf 309}, 323
                          (1989);
      D. Marvakov, A. L. Kuzemsky, and J. Vlahov, Phys. Lett. {\bf
       A105}, 431 (1984). %
\bibitem{lacroix}  C. Lacroix, J. Phys. F {\bf 11}, 2389 (1981).
 %
\bibitem{shoram} W. Shockley, J. Appl. Phys. {\bf 9}, 635 (1938); S.
             Ramo, Proc. IRE {\bf 27}, 584 (1939).
%
\bibitem{ssc93} V. Valtchinov, T. Ivanov, and L. T. Wille,
Solid State Commun. {\bf 89}, 637 (1994).
%
\bibitem{foxman93} E. Foxman, P. McEuen, U. Meirav, N. Wingreen, Y. Meir,
P. Belk, M. Kastner, and S. Wind, Phys. Rev {\bf B 47}, 10020 (1993).
%
\bibitem{averin91} D. V. Averin, and K. K. Likharev, in {\it Mesoscopic
Phenomena in Solids}, B. L. Altshuler, P. A. Lee, and R. A. Webb Eds.,
(North-Holland, Amsterdam 1991).
%
\bibitem{fren89} W. R. Frensley, Rev. Mod. Phys. {\bf 62}, 745 (1989).
\end{references}
\end{document}